\documentclass[aps,pra,oneside,twocolumn,showpacs]{revtex4-1}
\usepackage{graphicx}
\usepackage{epsfig}
\usepackage{amssymb}
\usepackage{multirow}
\usepackage{booktabs}

\usepackage{dcolumn}
\usepackage{bm}
\newcommand{\ignore}[1]{}

\usepackage{color}
\usepackage{amsmath}

\begin{document}
\def\e{\mathcal{E}}

\title{Structure and Symmetry in Coherent Perfect Polarization Rotation}

\author{Michael Crescimanno,$^*$ Chuanhong Zhou, James H. Andrews, Michael A. Baker}
\affiliation{Department of Physics and Astronomy, Youngstown State
University, Youngstown, OH 44555-2001, USA}
\email{dcphtn@gmail.com}

\date{\today}

\begin{abstract}
Theoretical investigations of different routes to coherent perfect polarization rotation illustrate its phenomenological  connection with coherent perfect absorption. Studying systems with broken parity, layering, combined Faraday rotation and optical activity, or a rotator-loaded optical cavity highlights their similarity and suggests new approaches to improving and miniaturizing optical devices. 

\end{abstract}


\maketitle
\section{INTRODUCTION} 

\begin{figure}[b]
\includegraphics[width=\linewidth]{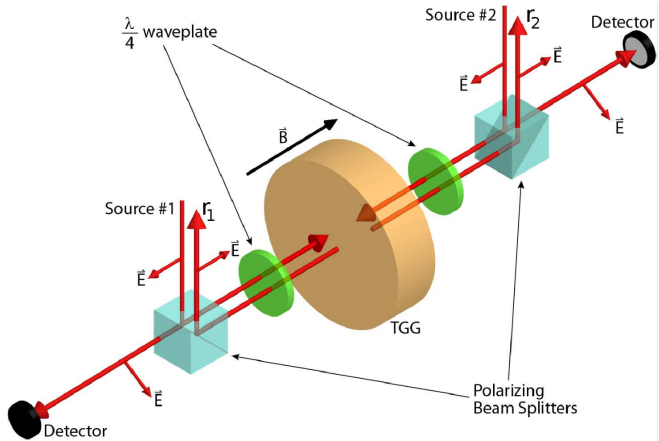}
\caption{Basic schematic of two-port CPR device. Not shown are 
attenuators and delay plates that balance the input field's 
amplitude and phase. When the CPR resonance condition is achieved, the 
reflected light $r_1$ and $r_2$ vanishes.}
\label{CPRbasic}
\end{figure}

Coherent perfect polarization   
rotation (or CPR)~\cite{CPR1}  is a conservative, reversible 
example of a multiport, maximally efficient, optical mode conversion process. 
As such
it shares phenomenological correspondences with 
the coherent perfect absorber
(antilaser or CPA)~\cite{chong10.01,wan11.01} which has been
well studied~\cite{longhi10.01, chong11.01, longhi11.01,lin11.01}. 
While many optical devices such as laser wavelength locks, 
field sensors, optical isolators, and modulators 
are based on the non-reciprocal nature of 
Faraday rotation, one way to improve all of these devices 
is to  process all of the incident light coherently. CPR-based design is 
an intrinsically multi-(input)port approach that combines the 
non-reciprocal nature of 
the Faraday effect with interference to convert {\it all}  of the incident light
into its orthogonal polarization. An example of the basic two-port
CPR device is shown in Fig.~\ref{CPRbasic}. 

For the apparatus shown in Fig.~\ref{CPRbasic}, one still must tune the magnetic field
to specific values to achieve complete conversion of the polarization. The required  
field  is significantly below that of a single port rotator, however. For one example, the complete conversion of one polarization into the orthogonal polarization using an uncoated terbium gallium garnet (TGG) slab requires only 60\% of the field-length product needed for the same rotation in just transmitted light~\cite{CPR1}.  Achieving 
complete polarization conversion at a  lower field-length 
product is technologically
useful because it is precisely the seeming
``incompressibility'' of this product and 
the modest Verdet coefficients of commercially available optical materials that 
pose a major obstacle to the diffusion of single port designs into integrated
optical assemblies and low cost devices. The primary motivation for 
the work reported here is to quantify in typical 
one-dimensional optical geometries
how CPR-based optical design significantly lowers the  
field-length product. We characterize the reduction of 
the threshold field-length product at which coherent perfect processes first
occur consequent to specific design choices 
in optical dispersion, structural dispersion, 
broken parity, and localization. 

A second motivation for this study is to reveal general principles
common among coherent perfect processes. Enlarging the context for 
these phenomena builds intuition useful for finding routes
to improving optical devices. We do this primarily 
by comparing and contrasting CPR and CPA. 
Note that in both CPR and CPA,  a tunable time-odd optical process 
(magneto-optical rotation for CPR versus absorption in CPA) 
is combined with multiport wave interference
to achieve perfectly efficient mode conversion
(to orthogonal polarization in CPR and to electronic excitation in CPA). 
Among other commonalities between CPA and CPR are their 
critical dependence on the relative optical phases among 
the input lightfields. As in the single port case of 
critical coupling, CPA and CPR both require a particular
(hereafter ``threshold'') magnitude for the time-odd 
process. In both CPR and CPA the conversion efficiency has a 
resonance-like structure. Also, for a fixed wavelength
going both above or below threshold makes complete conversion impossible. 
Because this ``resonance''
is not associated with a particular decay timescale, coherent perfect 
resonances are intrinsically zero-width. 

Understanding of coherent perfect absorption (CPA) phenomena in diverse optical systems has advanced steadily. Theory relates CPA states to self-dual spectral singularities~\cite{chong10.01,mosta} of the S-matrix. The CPA threshold's dependence on the depth of the sample is well understood theoretically, and this dependence has been modeled in complex (but still linear) media such as gold-silica composites and other plasmonic systems~\cite{agarwal,agarwal2}, and nonlinear media~\cite{longhi_nonlin}. Both ordinary and ${\cal P}{\cal T}$-symmetric systems elicit a diverse set of CPA phenomena with those most relevant to CPA thresholds including gratings~\cite{grating1,PTgrating,metasurfaceCPA,metasurfaceCPA2}, surface plasmonic polaritons~\cite{metasurfaceCPA3}, photonic crystals~\cite{PT_PhC}, near-zero-$\epsilon$ materials~\cite{eps_zero}, cavities with absorbers~\cite{cavity_abs}, controlled disorder or other spatial ordering~\cite{disorderedCPA,spatial_variation}, and very thin absorptive layers~\cite{absorb_thin}. Some of these ideas are also being explored for technological uses including all optical switching~\cite{CPAswitch,fastswitch,yet_another_CPA_switch,yep_another_one,diffractive_switch} and CPA enhancement of photoluminescence~\cite{PLenhance}. 

After fixing notation and giving
a brief review of the basic phenomena, 
Section III explores CPR and contrasts it with CPA 
in different optical environments, highlighting the roles played by dispersion, parity, and
transport of mixed symmetry type.  In Section IIIA we focus on coherent 
perfect phenomena in model multilayer systems, followed in Section IIIB by breaking parity two 
different ways (first softly with randomness and then explicitly with
trinary multilayers). Thematically up to that point, 
one sees a direct correspondence between the layering effects on CPR and CPA states 
and thresholds.  Subtle differences between the two are discussed
in one archetypal example combining CPR and optical rotation in Section 
IIIC. A brief conclusion highlights new directions prompted by this study. 

\section{NOTATION AND PRELIMINARIES} 

We use matrices to represent linear transport and 
work in the basis where the local field (complex) amplitudes 
for light traveling along the ${\hat z}$-axis are
${\vec v} = (E_x, H_y, E_y, -H_x)$. In terms of the individual polarization
and motional states,  we use ${\vec e}_R = (E_x, H_y) = (1,1)$ for a right-moving wave and ${\vec e}_L = (-1,1)$ for a left-moving one. 
Throughout this paper we restrict ourselves to 
materials without linear birefringence (in contrast with ~\cite{wang1})
The $O(2)$ symmetry about the axial
direction implies for the transport 
${\vec v}_{i+1} = {\cal M}_i {\vec v}_i$ that the 4$\times$4 ${\cal M}$ can be written 
(in this basis) in terms of the 2$\times$2's $M$ and $C$  as 
${\cal M} =
\left( \begin{array}{cc}
M & C \\
-C & M \end{array} \right)$,
where $C$ is only nonzero for transport that mixes the polarization states. 

For dielectrics (also the only case we consider below), the 
matrix $M$ is proportional to the familiar 2$\times$2 transfer matrix 
for the individual polarizations. For example, for a unit intensity wave 
incident from the left, in steady state, the field amplitudes at the 
surface are   
${\vec e}_{\rm in} = (1,1) + r(-1,1)$, where $r$ is the reflected amplitude. 
The outgoing field amplitude to the right of the system is given via 
${\vec e}_{\rm out} = t(1,1) = M{\vec e}_{\rm in}$, 
where $t$ is the transmission amplitude. In this basis, for a 
purely dielectric material of thickness $L$, index $n$, 
\begin{equation} 
M =
\left[ \begin{array}{cc}
\cos\delta & {\frac{i}{n}}\sin\delta \\
in\sin\delta & \cos\delta\\
\end{array} \right], 
\label{Mdielectric_2by2}
\end{equation}
where $\delta = nk_0L$ and $k_0$ is the vacuum wavenumber.
Note that $det(M)=1$ always, but $M_{11}$ and $M_{22}$ are only 
equal in systems that have overall spatial parity symmetry. 
We identify the real part of the index $n$ with refraction and its positive/negative imaginary part with absorption/gain. 

Analytically for a slab dielectric Faraday rotator
the $M$ and $C$ parts of the ${\cal M}$ in our field 
basis are~\cite{CPR1,kato03.01}
\begin{equation}
M =  {\frac{1}{2}} \left[ \begin{array}{cc}
C_1+C_2 & i(S_1/n_1+S_2/n_2)  \\
i(n_1S_1+n_2S_2)  &  C_1 + C_2 \end{array} \right]
\label{Mdielectric}
\end{equation}
and
\begin{equation}
C = {\frac {1}{2}} \left[ \begin{array}{cc}
i(C_1-C_2) & -(S_1/n_1-S_2/n_2)  \\
-(n_1S_1-n_2S_2)  &  i(C_1 - C_2) \end{array} \right] \, ,
\label{MnCdielectric}
\end{equation}
where $C_{1,2}$ ($S_{1,2}$) refer to the cosine (sine) of
$\delta_{1,2} = n_{1,2} k_0 L$ in which the $n_1,n_2$ are the indices
of refraction
of the left- and right- circular polarization in the slab,
$k_0$ refers to the vacuum wavevector, and $L$ is the thickness of
the slab. For a dielectric slab in an
external magnetic field pointing along the direction of propagation,
$\Delta n=n_1-n_2 \propto VB$, the product of the Verdet and the magnetic
field. Note that the resulting 4$\times$4 matrix 
${\cal M}$ is quite different from one 
representing optical activity (a time-even rotation process)  
which has the form $M= \cos\alpha M_0$ and $C=\sin\alpha M_0$, where $\alpha$ is proportional to the density of 
chiral centers in the slab and $M_0$ is the usual 2$\times$2 transfer matrix 
given by Eq.~(\ref{Mdielectric_2by2}). 
Because CPR is a reversible optical process we require constant local power flux
throughout in steady state. This condition thus requires the $n$'s and the $\alpha$ to be real throughout for both the time-even and time-odd rotation 
processes we consider below. 

For a single polarization whose linear transport is given 
entirely in terms of a net 2$\times$2 transfer matrix $M$  in the basis 
described above and used throughout, 
the CPA state is reached when the condition  
$(1,1)M(1,1)^t = M_{11} + M_{12} + M_{21} + M_{22} = 0$ 
is satisfied. For a general 
2$\times$2 matrix, 
this condition combined with the determinant indicates that CPA 
implies four real conditions for four complex numbers. A remaining freedom 
of optical field  (amplitude and phase) then implies that 
CPA  requires, at minimum,  tuning two dimensionless
experimental parameters, typically, the ratio $L/\lambda$ and the absorptive
index $Im(n)$. 

It is also straightforward to find the condition associated with 
CPR resonances using the
4$\times$4 basis. For fields incident from the left, take
${\vec v}_l = (1,1,-l, l)$, where $l$ is the amplitude of the reflected,
rotated wave. On the right, take ${\vec v}_r =
(-d, d,s,s)$;  this configuration thus consists of  incoming
fields of one polarization and  outgoing fields of the orthogonal
polarization only, the CPR state.
In analogy with the CPA state, these boundary conditions
lead to a condition on the
size, wavelength and rotary power of the system.
For CPR resonance in uniaxial systems
with the 4$\times$4  form of ${\cal M}$ as described earlier, we require
\begin{equation}
M \left( \begin{array}{c} 1 \\ 1 \end{array} \right) + C \left( \begin{array}{c} -1 \\
1  \end{array} \right) l
=
 \left( {\begin{array}{c}
-1 \\
1  \end{array}} \right) d
\label{CPRa}
\end{equation}
and
\begin{equation}
-C
\left( \begin{array}{c}
1 \\
1  \end{array} \right)
+ M \left( \begin{array}{c}
-1 \\
1  \end{array} \right) l
=
 \left( \begin{array}{c}
1 \\
1  \end{array} \right) s \, .
\label{CPRb}
\end{equation}
Counting conditions (four complex) for the three complex 
fields ($d,l,s$),  we see
that to achieve CPR by simultaneously 
solving Eqs.~(\ref{CPRa}) and (\ref{CPRb}) requires, at minimum, tuning 
two experimental parameters (here, generically, the ratio ${L/\lambda}$
and the circular birefringence $\Delta n=n_1-n_2$), which we note is 
analogous to the CPA case (where the parameters are $(L/\lambda)$ and the absorption coefficient).  
Eliminating the fields $d,l,s$,  we can write the CPR condition succinctly for 
a general ${\cal M}$ as 
a single complex condition $det(R)=0$,  where the 2$\times$2 matrix $R$ has the 
following elements:  
\begin{equation}
R_{11} = (-1,1)C^{-1}M\left( \begin{array}{c}
-1 \\
1  \end{array} \right) \,
\label{CPR_simpler1}
\end{equation} 
\begin{equation} 
R_{12}= -(-1,1)C^{-1} \left( \begin{array}{c}
1 \\
1  \end{array} \right) \,
\end{equation}
\begin{equation}
R_{21} = (1,1)[MC^{-1}M+C]\left( \begin{array}{c}
-1 \\
1  \end{array} \right)  \,
\end{equation}
\begin{equation}
R_{22} = -(1,1)MC^{-1}\left( \begin{array}{c}
1 \\
1  \end{array} \right)  \, .
\label{CPR_simpler4}
\end{equation}

We now summarize
CPR phenomenology in a series of optical systems
in order to build a deeper intuition about the CPR state and 
its connection to and contrast with CPA, with an eye towards its potential utility
in optical devices. 

\section{ CPR IN MODEL SYSTEMS} 

\subsection {CPR in layered binary systems} 

Studying CPR in multilayer interference films provides a 
straightforward comparison of CPR and CPA phenomena and their dependence on 
dispersion, both material and structural. For simplicity, consider
first a perfectly periodic multilayer composed of $N$ alternating layers of 
a material $A$ that is a dielectric 
with zero Verdet and a material $B$ that has a non-zero Verdet. 
We compare these systems to the CPA model system in which the bilayers
have one non-absorbing species ($A$) and the other absorbing ($B$). 
In all the model systems described here, only the $B$ species rotates (for CPR) 
or absorbs (for CPA). 
The species
have different indices of refraction in the absence of a magnetic field (for CPR) or absorption (for CPA) that we denote $n_A$ and $n_B$, creating 
an optical (reflection) bandgap. We denote these layered systems 
as $(AB)^N$, but here, to eliminate any spurious effect
from explicitly broken parity, we restrict our attention to 
parity symmetric layered systems 
formed by adding one terminal $A$ layer, that is, $(AB)^NA$. 

\begin{figure}[t]
\includegraphics[width=\linewidth]{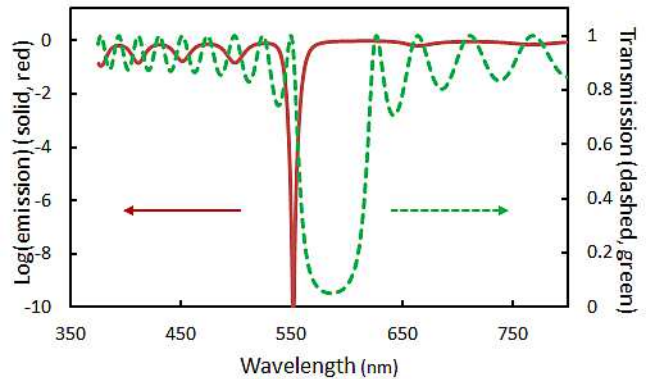}
\caption{(Color online) Typical CPR resonance in a 1-d photonic crystal. The green dashed  trace is the transmission trace and the red solid trace is the total output  light in the same polarization as the input polarization. 
This example is 31 
alternating layers (each 100 nm thick) 
of index 1.55 (non Faraday) and 1.38 (Faraday), 
corresponding to the first row in Table \ref{table1}~\cite{plastic}.
Clearly seen in the  dashed green trace is the reflection bandgap that extends from 550-625 nm. 
The CPR resonance is the pronounced reduction in the output light polarized along the input polarization for wavelengths near the short wavelength edge of the band.}
\label{foldedDFB_CPR}
\end{figure}

\begin{table}
\caption{CPR thresholds ($\Delta n$ values for the circular polarization propagation eigenstates in layer $B$) at the reflection band edge for the layered binary systems described in Fig. \ref{foldedDFB_CPR}. Throughout this paper the letters after the threshold values indicate the spatial symmetry ($O$ for odd, $E$ for even) of that CPR resonance's fields.} 
\label{table1} 
\begin{tabular}{ c c | c c c c } 
\hline
\hline
configuration &&& 31 layers && 33 layers \\
\hline
$n_A>n_B$  &&& .049(E) && .044(O) \\
$n_A<n_B$  &&& .090(E) && .080(O) \\ 
\hline
\hline
\end{tabular}
\end{table}

One finding of these simulations (see Fig.~\ref{foldedDFB_CPR})
is that for the lowest 
thresholds, layered films  with an 
odd number of bilayers ({\it e.g.}, 31 layers) had threshold CPR states of 
even parity and those with 
an even number of bilayers ({\it e.g.}, 33 layers) had threshold CPR states 
of odd parity. 
A simple explanation of this observation is given in 
the next subsection on consequences of parity symmetry (and parity breaking).

\begin{figure}
\includegraphics[width=\linewidth]{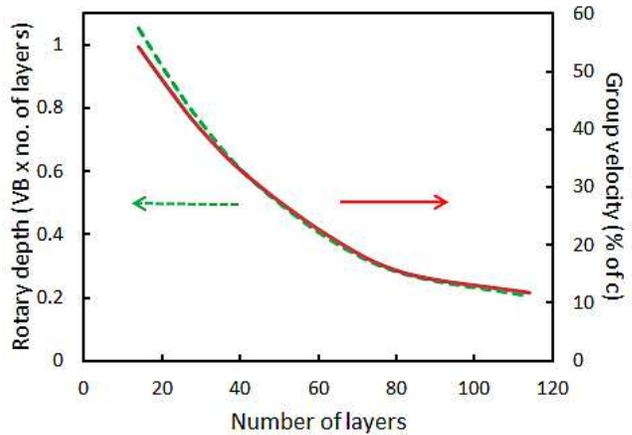}
\caption{(Color online) The correspondence between the group velocity delay and the CPR thresholds 
in layered media. Varying the number of layers only, we plot
the product of the CPR threshold Verdet-field ($VB$) 
product times the number of layers
versus the number of layers.  We also plot the 
group velocity minimum at the band edge versus the number of layers (right axis, solid red trace). }
\label{multilayer_effect}
\end{figure}

Note also that the 
wavelength at which the lowest CPR resonance occurs is
at a band edge. 
As is well known, across the reflection 
bandgap there is pronounced optical dispersion 
resulting in large increases
in the group velocity delay symmetrically 
at the band edges, and significant reductions in the delay in the middle of 
the band. (Ref.\cite{cresc12.01} is a recent relevant summary.) 
The reduction in the threshold for CPR/CPA with 
increases in group velocity delay at the band edge is most clearly seen by plotting 
the product of  
the threshold value of the rotary power of the $B$ layers times the number of layers versus 
the number of layers, as in Fig. \ref{multilayer_effect} (dashed green trace) 
which hews closely to a plot of the 
group velocity minima versus the number of layers (solid red trace). 

Changes in the group velocity delay
are, in and of themselves, not enough 
to explain the pattern of CPR (and CPA) resonances in these systems; the 
lowest threshold for the CPR/CPA resonances for $n_A>n_B$ occurs at the 
short wavelength side of the bandgap but occurs at the long wavelength 
edge of the bandgap for $n_A<n_B$. 
As the time reverse of CPA, actual 
lasing~\cite{Dowling}, indicates, the simplest way to explain this
difference is by apportioning the group velocity delay 
across the two species of the multilayer, 
and noting that only in the $B$ species is the light subject to polarization
rotation (CPR) or absorption/gain (CPA/lasing). Simulations of the local 
electric field of the light traversing the multilayer indicate that, off-band,
the apportionment of the total velocity delay
should follow the ratio of the indices. Near the 
long wavelength side of the band edge, however, the light's integrated electric energy
is greater in the larger index species, 
whereas the reverse occurs at the short wavelength edge of the 
bandgap~\cite{YSUreview}. Equating field energy to the probability that the light visits that 
species, and apportioning part of the total propagation time
(and thus the overall Faraday rotation) to each species in proportion to that 
probability, qualitatively explains both the wavelength and 
the threshold of the CPR/CPA resonances. 

Just as in CPA, there are other CPR states that arise at 
different wavelengths as one increases  $\Delta n$ beyond threshold. 
Note that for these simple binary  multilayers, the 
next-to-lowest CPR resonance typically 
occurs at the opposite band edge, as expected and formerly noted~\cite{grating1} for CPA. To summarize the results from this study of CPR in multilayers
in experimental terms, a 0.76 mm thick multilayer of 160 nm layers of each
Bismuth-substituted iron garnet (called ``BIG'', index of 
refraction of $\sim$ 2.3 at 633 nm, at which the Verdet is $\sim$ -7 x 10$^3$ rad/Tm ~\cite{kato03.01}) and ordinary SiO$_2$ glass would achieve CPR at 0.5 T, whereas
a slab of BIG alone of length 2.4 cm  would be needed at this field, indicating in concrete terms the substantial reductions in CPR thresholds 
associated with photonic bands. 

In practice with real multilayer systems, non-ideality typically moves CPR/CPA 
lowest threshold states from the band edge 
to defect states in the band gap itself. 
This is consonant with the experience in lasing where it is well 
documented that layer nonuniformity and other perturbations cause 
lasing to occur first through defect states typically located within 
the band gap itself. The defect states still correspond to maxima of the
group velocity delay~\cite{wu09.01}.  (See the discussion of parity breaking in subsection B below.)

\begin{figure}[t]
\includegraphics[width=\linewidth]{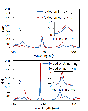}
\caption{(Color online) Transmission gain versus wavelength for the folded structures analogous to those in Table \ref{table2}, where instead of rotation in the $B$-layers a complex index of refraction is used to represent optical gain. In this case the optical band stretched from 450 nm to 520 nm, and the prominent fold defect state appears near the center of the band. (Figure adapted from Ref.~\cite{DFBlaser} with permission of The Optical Society of America.)} 
\label{foldedDFBgain}
\end{figure}

It is illustrative to explicitly compare thresholds for 
lasing/CPA and CPR in simple multilayer systems with 
deliberate structural defects, such as the ``phase-slip'' (sometimes called
``folded'') distributed feedback (DFB) systems~\cite{YSUreview,yablanovich}. 
Here we compare the multilayers $(AB)^N(BA)^N$ and $(BA)^N(AB)^N$, where the 
time-odd process (either Faraday rotation in the case of CPR or absorption/gain
in the case of CPA/lasing) is again only in the $B$-layers. 
  Table \ref{table2} gives  calculated threshold $\Delta n$ values for CPR for four configurations of simply-folded symmetric systems, 
which agree qualitatively with the corresponding results for CPA/lasing 
summarized in Fig. \ref{foldedDFBgain} (adapted from Ref.~\cite{DFBlaser}). 
For example, controlling for overall gain, the folded DFB structure with the lowest lasing 
threshold (as inferred from the largest gain in the figure) is that which has the gain medium 
in the low index material and is folded on the low index material. 
This result agrees with our simulations of the CPR threshold 
as shown in Table \ref{table2} (folded on $B$, $n_A>n_B$).

\begin{table}
\caption{The CPR threshold values of $\Delta n$ for ``folded'' layered systems comprised of 52 total layers. Every entry in the table is for a CPR resonance occurring on the  defect state inside the reflection band.  The lowest CPR threshold occurs with even parity when  rotation occurs in the lower index material and the fold is on that low index material. The CPR threshold ordering in the chart is in one-to-one agreement with that of lasing thresholds in these ``folded'' DFB systems reproduced in 
Fig.~\ref{foldedDFBgain}.} 
\label{table2} 
\begin{tabular}{ c c | c   c  c  c } 
\hline
\hline
configuration &&& Fold on $A$ && Fold on $B$ \\
\hline
 $n_A>n_B$  &&& .032(O) && .013(E) \\
 $n_A<n_B$ &&& .017(E) && .028(O) \\ 
\hline
\hline
\end{tabular}

\end{table}


For contrast we conclude this section with a case in which dispersion, but 
not field placement, is important: the loaded optical 
cavity as a layered optical system. Consider 
a dielectric Faraday rotator inside an optical cavity composed of transversely 
isotropic perfectly thin mirrors of reflectivity amplitude $r$ (so that the 
reflectivity is $R = |r|^2$).  The mirrors are represented by the transfer
matrix ${\cal M}_{r} = \left[ \begin{array}{cc}
M_r & 0  \\
0 & M_r   \end{array} \right]$ where the 2$\times$2  
matrices $M_r$ for the simple case of completely non-absorbing mirrors
are given in our 'field' 
basis as
$M_r = {{1}\over{\sqrt{1-|r|^2}}}\left[ \begin{array}{cc}
1 & i|r|  \\
-i|r| & 1  \end{array} \right]$. 
Algebra indicates that all effects of the cavity reflectivity
modify the conditions for CPR via a single parameter, 
$\gamma = {{2|r|}\over{1+|r|^2}}$. 
One finds for this loaded cavity configuration
(mirror-rotator-mirror) that the CPR condition becomes ~(compare  the 
$r \rightarrow 0$ limit with Eq.(14) in Ref.~\cite{CPR1}):
\begin{equation} 
(n_1+{{1}\over{n_1}}) S_1C_2-(n_2+{{1}\over{n_2}})S_2C_1+\gamma
\bigg({{n_1}\over{n_2}}-{{n_2}\over{n_1}}\biggr)S_1S_2 = 
\nonumber
\end{equation}
\begin{equation} 
\pm\biggr[(n_1-{{1}\over{n_1}})S_1-(n_2-{{1}\over{n_2}})S_2 + 2\gamma(C_2-C_1)\biggr] \, ,
\label{CPR_cavity} 
\end{equation}
where, as before,  $n_{1,2}= n_0\pm \Delta n/2$ and $k_0 \Delta n = 2VB$.

To compare this result with CPA, it is straightforward to show that 
the lowest CPA resonance threshold for a cavity loaded with a lossy dielectric
modeled as a complex index $n$ is given by the solution of  
(compare with the $r \rightarrow 0$ limit of Eq.~(7) of Ref.~\cite{chong10.01} also reproduced below in Eq.~(\ref{CPA_basic}) for completeness):  
\begin{equation} 
e^{i2nk_0L} = 
{{(n-1)^2-(n^2+1){{2R}\over{1+R}} +i\gamma (n^2-1)}\over
{(n+1)^2-(n^2+1){{2R}\over{1+R}} +i\gamma (n^2-1)}}
   \, . 
\label{CPA_cavity} 
\end{equation}

\begin{figure}[t]
\includegraphics[width=\linewidth]{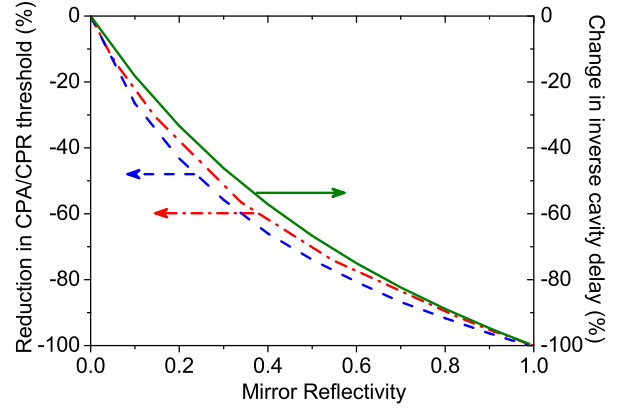}
\caption{(Color online) Lowest CPR (red, dot-dash trace) and CPA (blue, dashed trace) resonance thresholds in a loaded cavity decrease as one increases the finesse of the mirrors, going to zero with the inverse of the group velocity delay (green, solid). } 
\label{CPR_cav}
\end{figure}

In Fig. \ref{CPR_cav}, we have 
used Eq.~(\ref{CPR_cavity}) for $n_0 = 2.0$ and $k_0L \sim 820$ 
to plot the fractional 
reduction in the lowest CPR resonance threshold ($\Delta n$) as 
a function of the reflectivity, $R$. 
The graph shows strong similarity to the inverse of the 
time spent in the cavity ({\it i.e.} the fractional reduction in the group velocity), as expected, and also corresponds with  the reduction in 
the CPA threshold of an absorber-loaded cavity shown in the graph. 
Not included in Fig.~\ref{CPR_cav}, 
we have also analyzed
a realistic ({\it e.g.} complex dielectric) gold mirrored cavity at 
780 nm and find qualitatively the same behavior as in Fig. ~\ref{CPR_cav} 
with increasing 
gold layer thickness. In that study there are no 780 nm CPA states 
from tuning the loss in the dielectric slab inside the cavity if the gold
layer thickness exceeds $~$35 nm (corresponding to an $R$ of about 85\% in each 
mirror) because at that depth the absorption in the gold itself is  
above the CPA threshold. 

\subsection{CPR with explicitly broken parity} 

\begin{figure}[t]
\includegraphics[width=200pt]{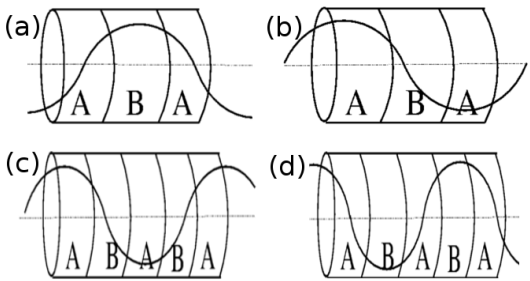}
\caption{ Parity and CPA/CPR. For an odd number of bilayers, we see that
(a) the parity even fields have a 
maximum on the $B$ species, whereas  (b), the parity odd case, the field on the $B$ layer
 is significantly smaller.  For an even number of bilayers, however, the situation is reversed so that (c) parity even fields do not have their maxima on the $B$ layers, but (d) parity odd fields do. 
} 
\label{cylinders}
\end{figure}

In a parity symmetric absorbing structure, the fields of all CPA states must also be of definite parity, even or odd. 
These two possibilities 
generally occur at different absorption thresholds. 
Since we have already 
discussed layered optical systems, one particularly intuitive 
way to understand this difference is shown in Fig. \ref{cylinders}, where one of the species  ($B$) is absorbing (or rotating in the CPR case) and the other species ($A$) is not. 
For wavelengths nearly four times the layer thickness (near the band edge)
as in the example shown in Fig.~\ref{cylinders},
the parity even case has larger field overlap 
on the absorber/rotator (species $B$) than the parity odd field for an 
odd number of bilayers (in this case one bilayer), thus the former will have a lower CPA threshold (compare with Table~\ref{table1}).

When a rotator (or absorber) is not parity symmetric, there are still 
CPR (CPA) states, but the state's fields will not be of definite parity. 
To illustrate the effect of parity breaking on CPR and 
its comparison with CPA (See Refs.~\cite{chong11.01,brokenP_CPA}), in this section we 
consider two examples of parity broken systems:
 (i) an $(AB)^NA$ multilayer,  but with layer-to-layer 
thickness variations, and (ii) 
a trinary regular layered system of the
type $(ABC)^N$ (in both cases only $B$ is rotary (CPR) or absorptive (CPA)).

\begin{figure}[t]
\includegraphics[width=\linewidth]{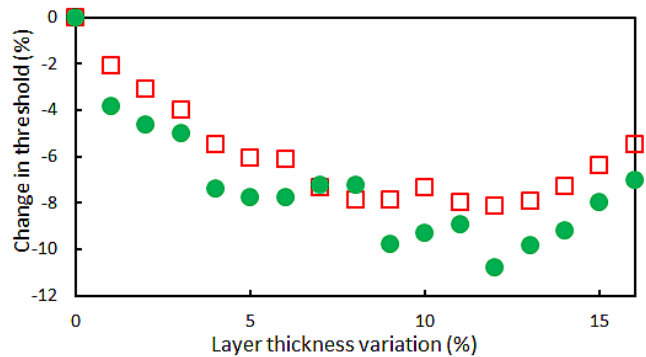}
\caption{(Color online) Percentage reduction of CPR (closed circles, green) and CPA (open squares, red) in the thresholds for the lowest CPR/CPA state, which, as described in the text, occur near the band edge. For this example, a particular random layer thickness variation map for the 65 layer multilayer with $n_B>n_A$ (only species $B$ is Faraday (CPR case) or absorptive (CPA case)) is programmed into the simulation and increased across the horizontal axis.}  
\label{localization}
\end{figure}

As one introduces layer thickness variations into the $(AB)^NA$ structures
discussed in the preceding subsection, formerly localized reflection band states 
mix with extended states whereas some formerly extended states become localized~\cite{wu09.01}. 
Initially, weak localization increases the group velocity delay and thus reduces the CPA/CPR threshold for some states near the band edge (see Fig.~\ref{localization} for one example). As the localization length shrinks further with increasing layer thickness variations,  random scattering reduces the coherent band edge reflections that were responsible for the increase in the group velocity delay in the first place.
 As the level of randomness is increased, the lowest resonant CPR/CPA state's wavelength at threshold moves into what was previously the reflection band.  
Note also that adding layer thickness 
randomness explicitly breaks the original parity 
symmetry of the system. As a consequence, at finite randomness
in the CPA case, the amplitude ratio of the input fields is no longer $\pm 1$. 

The consequence of parity breaking through broken structural  symmetry in CPR is different from that of CPA~\cite{paritynote}.
Solving Eqs.~(\ref{CPRa}),(\ref{CPRb}) for the amplitude ratio of the incident fields, $l$, indicates 
\begin{equation} 
l={ {(M_{11} + M_{12}+M_{21}+M_{22})} \over { (C_{11}+C_{21}-C_{12}-C_{22})}}
\nonumber
\end{equation}
\begin{equation} 
= {{(C_{11}+C_{21}+C_{12}+C_{22})} \over {(M_{22}+M_{12}-M_{11}-M_{21})}}
\label{CPR_eq_for_l}
\end{equation} 
on the CPR state. 
For any optical system composed of sections without birefringence or optical activity, it was shown in Ref.~\cite{CPR1} that the 4$\times$4 ${\cal M}$ has underlying 2$\times$2 matrices $M$ and $C$ with the $M$ being time-even and of the form $\left[ \begin{array}{cc}
{\cal R}  & {\cal I}  \\
{\cal I}  & {\cal R} \\
\end{array} \right]$
and the $C$ being time-odd and of the form
$\left[ \begin{array}{cc}
{\cal I}  & {\cal R}  \\
{\cal R}  & {\cal I} \\
\end{array} \right]$, where ${\cal R}$ means a purely real quantity and 
${\cal I}$ a purely imaginary one. This matrix structure in the formulas for $l$ gives
$|l|=1$, which in turn forces all the optical fields to have the 
same intensity for a CPR state in any system. 
The broken parity in the CPR state is instead manifest by the appearance of mismatched phases (not shown)
between the input fields. 

\begin{table}
\caption{Thresholds for $(ABC)^N$ layered systems where the indices of refraction are chosen from the list  \{1.55, 1.51, 1.38\} and the total number of layers is 45 (each layer is 100 nm thick.)} 
\label{table3}
\begin{tabular}{ c  c  c  c  c  c  c  } 
\hline
\hline
TYPE && configuration && threshold && amplitude ratio \\  
\hline
CPR && $n_A>n_C>n_B$  && .0465 && 1  \\ 
CPR && $n_A>n_B>n_C$  && .0652 && 1  \\ 
CPR && $n_B>n_A>n_C$  && .0752 && 1  \\ 
CPA && $n_A>n_C>n_B$  && .0528 && 1.44  \\ 
CPA && $n_A>n_B>n_C$  && .092 && .215 \\ 
CPA && $n_B>n_A>n_C$  && .087 && 1.52  \\ 
\hline
\hline
\end{tabular}
\end{table}

Explicit parity breaking via structural asymmetry is also evident in the trinary films, $(ABC)^N$, again, in which only $B$ is rotary (CPR) or absorptive (CPA). As an example, Table \ref{table3} shows a comparison of CPR and CPA lowest resonance thresholds near the band edges  of the very first reflection band of   trinary films. 
Thus, even for perfectly ordered trinary films, it is the phase mismatch between the left and right input fields that varies universally, while the amplitude ratios only vary for the CPA case. Note also that the threshold values for the cases $n_A>n_C>n_B$ and  $n_B>n_A>n_C$ are ordered the same in both CPR and CPA. In particular, for each of these cases in Table ~\ref{table3}, the CPR/CPA state forms at the appropriate band edge as discussed in the previous section.
The spectral location of the CPR/CPA state in the intermediate case $n_A>n_B>n_C$ depends on the indices' values. 

Two additional facts of interest emerge from these simulations. As one might expect, the intensity ratios are more varied for the trinary films (explicit parity breaking) than for the random $(AB)^NA$ layered system (which breaks parity more softly) studied here to only 15\% layer thickness variation. Also, for the  case of random $(AB)^NA$ layered systems, the variation in the phase is much larger in the CPR case than in the CPA case. Note in this regard  that the CPR state forces the intensity ratio to remain unity, whereas for CPA both the amplitude ratio and the phase adjust to stay resonant in a parity broken system. 


\subsection{Combined Faraday rotation and optical activity} 

\begin{figure}[t]
\includegraphics[width=\linewidth]{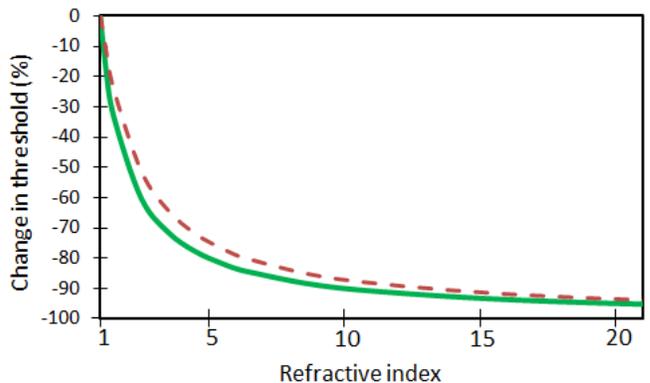}
\caption{The CPR threshold reduction (red, dashed) is monotonic to zero as one increases the index of refraction of the slab,  following the same relation as in the CPA case ($\sim 1/n$ at large $n$ as described in the text, here as a green, solid line). } 
\label{fig_CPR_index}
\end{figure}

\begin{figure}[t]
\includegraphics[width=\linewidth]{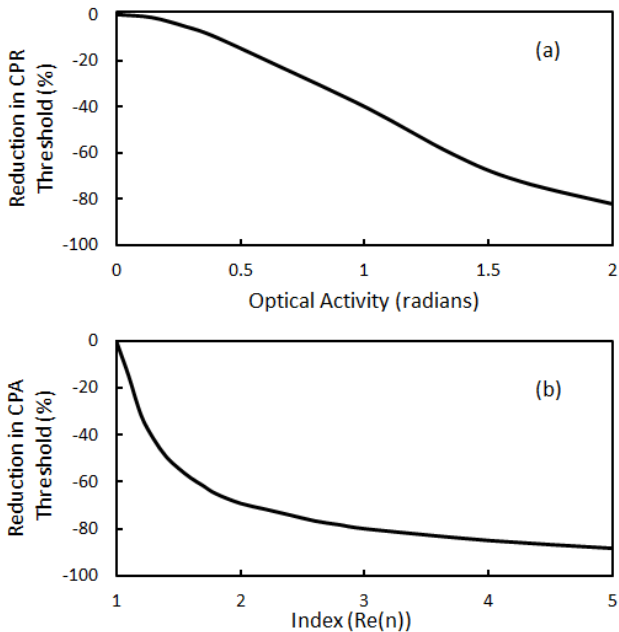}
\caption{Lowest CPR and CPA resonance thresholds  decrease monotonically to zero
as one increases the time-even part of the transport. (a) Reduction in CPR threshold for a Faraday (time odd) constituent sandwiched between two optically active slabs as a function of total optical activity of the time even part standing alone (in radians).  (b) Reduction in CPA threshold as a function of the real part of the material's refractive index. } 
\label{CPR_OFOgraph}
\end{figure}

To highlight the time-reversal symmetries underlying CPR and CPA, we now 
address the effect that the time-even part of the transport has on the 
CPR/CPA threshold. In the original derivation of the CPR effect~\cite{CPR1}
in a simple slab dielectric, 
increasing the index of refraction of the material reduces 
the CPR threshold, as shown 
graphically in Fig. \ref{fig_CPR_index} using the formula in Ref.~\cite{CPR1}.
This is also the case with the CPA
threshold, which (see Eq.(7) of Ref.~\cite{chong10.01}) for an absorbing 
slab dielectric of index $n = n_0+in_{abs}$ and length $L$ is:
\begin{equation}
e^{ink_0L} = \pm{{(n-1)}\over{(n+1)}}.
\label{CPA_basic}
\end{equation}
In the large $n_0$ limit, because the log is vanishing as $\sim 1/n_0^2$, the threshold $n_{abs}$ must decrease as $\sim 1/n_0$ at large $n_0$. A graph of this 
reduction of CPA in a bulk absorber from Eq. ~(\ref{CPA_basic}) is included 
in Fig. \ref{CPR_OFOgraph}b. We note in passing that this reduction is what one would expect for the single transit time reduction and not that associated with the etaloning as was the case for the optical cavity-assisted reduction in the thresholds. 

In CPA, the index of 
refraction real and imaginary parts can be considered as the time-even and time-odd contributions
to the transport. The analogous processes for the transport of the polarization 
are optical activity (time-even) and Faraday rotation (time-odd). 
Recall that one cannot achieve CPR with optical activity 
alone, but the question we would like to address is how the presence of 
optical activity in a system modifies the threshold Faraday rotation needed
for CPR. 

Consider
a system with both of these processes operating. Instead of a single bulk 
piece, for simplicity 
we analyze a three-layer system composed of two optically active 
 blocks
with a Faraday rotator in between. (See discussion below Eq.~(\ref{Mdielectric_2by2}) for the matrix representation of optical activity.)
It is then straightforward to 
identify the CPR state in this system, 
again in terms of the equation $det(R)=0$, where
the matrix elements of the 2$\times$2 complex matrix $R$ are as in 
Eqs.~(\ref{CPR_simpler1})-(\ref{CPR_simpler4}),
but where we make the substitutions for $M$ and $C$ via; 
\begin{equation} 
 \left( \begin{array}{c} M \\ C \end{array} \right) = 
\left( \begin{array}{cc}
\cos 2\alpha & -\sin 2\alpha \\
\sin 2\alpha & \cos 2\alpha \end{array} \right)
 \left( \begin{array}{c} M_0MM_0 \\ M_0CM_0 \end{array} \right) \, ,
\label{CPR_OFO} 
\end{equation} 
where $M_0$ is given by Eq.~(\ref{Mdielectric_2by2}) for the optically active blocks (with chiral density proportional to $\alpha$) and the $M$ and $C$ 
on the RHS of Eq.~(\ref{CPR_OFO}) are given by Eq.~(\ref{Mdielectric}) and Eq.~(\ref{MnCdielectric}), 
respectively, for the Faraday block.
 Keeping the indices and length the same, but changing only the optical 
activity, we can determine the location of the CPR state 
(see Fig. \ref{CPR_OFOgraph}(a)).
 We see that,  as in the decrease of the 
CPA threshold with  increasing real part of the refractive index $n_0$, the Faraday rotation
needed to achieve CPR resonance decreases monotonically as one increases 
the optical rotation in the adjoining slabs. 
We note that this reduction continues with increasing  optical activity beyond the
value  at which the optical rotary part of the assembly 
by itself would rotate a single input  ray to its orthogonal polarization (rotation by $\pi/2$)
 upon exiting in transmission. This result is true for both
positive and negative Verdet irrespective of the handedness of the optical activity; the trace in Fig. \ref{CPR_OFOgraph}(a) is 
symmetric about zero optical activity. Both of these (CPR and CPA) 
curves asymptote to zero threshold. 
This shows that increasing the time-even part of 
an optical process reduces the time-odd 
threshold for achieving CPR/CPA, and is expected to be useful for reducing
the size, complexity and cost of  devices based on CPR or CPA, for example by 
reducing the required magnetic field.

\section{Conclusions} 
CPR and CPA are phenomenologically congruent in how their thresholds depend on the system's symmetry, composition and geometry. As both are coherent perfect processes, this congruence follows from the underlying commonality they share through wave interference and discrete symmetry. Furthermore, this study reveals potential design routes to decrease  the size and/or magnetic field requirements for achieving CPR. For example, as detailed above, multilayering the rotating species can yield a 30-fold reduction in the naive length-field product. Similarly, even a poor optical cavity with just 60\% reflective mirrors reduces the CPR threshold length-field product by nearly 80\%. By layering with suitable optically active materials, high index materials, tertiary layered systems and layered systems with small layer thickness variations in the  stack, we have shown further reduction in the naive length-field product is achievable in CPR-based devices. 


\acknowledgments
The authors are grateful to the National Science Foundation for
financial support under grant number ECCS-1360725 and for
financial support from the Science and Technology Center for
Layered Polymeric Systems under grant number DMR
0423914.




\begin{thebibliography}{11}
\bibitem{plastic}
These somewhat arbitrary values were chosen for their utility in multilayer CPA/lasing modeling and to be consistent with previous studies by the authors in Ref.~\cite{DFBlaser} and do not affect the conclusions drawn here.
\bibitem{paritynote}
In CPR,  parity is not broken explicitly by the magnetic field (a psuedo-vector) itself.
\end{thebibliography}
\end{document}